\documentclass[aps,prd,11pt,twoside,tightenlines,nofootinbib,showpacs,preprint,superscriptaddress]{revtex4-1}
\usepackage[colorlinks=true]{hyperref}
\usepackage{xcolor}
\hypersetup{colorlinks=true, citecolor=blue, urlcolor=blue, linkcolor=blue}
\usepackage{amsmath,amssymb}
\usepackage[final]{graphicx}
\usepackage{subcaption} 
\usepackage{float}

\newcommand{\gev}{\textrm{ GeV}}

\begin{document}
\title{$a_0(1710)$-$f_0(1710)$ mixing effect in the $D_{s}^{+} \rightarrow K_S^{0} K_S^{0} \pi^{+}$ decay}

\author{Yu-Wen Peng}
\affiliation{School of Physics, Central South University, Changsha 410083, China}

\author{Wei Liang}
\email{212201014@csu.edu.cn}
\affiliation{School of Physics, Central South University, Changsha 410083, China}

\author{Xiaonu  Xiong}
\affiliation{School of Physics, Central South University, Changsha 410083, China}

\author{Chu-Wen Xiao}
\email{xiaochw@gxnu.edu.cn}
\affiliation{Department of Physics, Guangxi Normal University, Guilin 541004, China}
\affiliation{Guangxi Key Laboratory of Nuclear Physics and Technology, Guangxi Normal University, Guilin 541004, China}
\affiliation{School of Physics, Central South University, Changsha 410083, China}

\date{\today}

\begin{abstract}

With the measurements of the decay $D^+_s \rightarrow K^0_S K^0_S \pi^+$ by the BESIII Collaboration, we investigate this three-body weak decay via the chiral unitary approach for the final state interaction, where the resonances $S(980)$ and $S(1710)$ are dynamically reproduced with the interaction of eleven coupled channels, and the $W$-external and -internal emission mechanisms are considered at the quark level. Besides, we also take into account the contribution from the $P$-wave resonance $K^*(892)^+$ and  make a combined fit of the $K^0_S K^0_S$ and $K^0_S \pi^+$ invariant mass spectra measured by the BESIII Collaboration. The fitted results show that the enhancement around 1.7 GeV in $K^0_S K^0_S$ mass spectrum is overlapped with two visible peaks, indicating the mixing signal originated from the resonances $a_0(1710)$ and $f_0(1710)$ due to their different poles (masses). Thus, the decay $D^+_s \rightarrow K^0_S K^0_S \pi^+$ is helpful to reveal their molecular nature with the mixing signal, which can be more precisely measured in the future.

\end{abstract}

\maketitle

\section{Introduction}

Due to the confinement and non-perturbative properties of quantum chromodynamics, the internal structure of the hadrons is a debated issue in the particle physics. Especially the properties of some resonances found in the experiments are challenging to be explained using the conventional quark model, such as $f_0(500)$, $f_0(980)$, $a_0(980)$, $\Lambda(1405)$, and so on. 
Note that the states $f_0(980)$ and $a_0(980)$ were found more than 50 years ago~\cite{Astier:1967zz,Ammar:1969vy,Defoix:1969qx}, of which the properties were under debate for a long time~\cite{Godfrey:1985xj,Morgan:1990kw,Jaffe:1976ig,Achasov:1980tb,Au:1986mq,Flatte:1976xu,Weinstein:1982gc,Zou:1994ea,Janssen:1994wn,Tornqvist:1995ay,Oller:1997ti,Locher:1997gr,Oller:1998hw,Oller:1997ng}. 
Both of them are generally assumed to be the $K\bar{K}$ molecule~\cite{Weinstein:1982gc,Zou:1994ea,Janssen:1994wn,Tornqvist:1995ay,Oller:1997ti,Locher:1997gr,Oller:1998hw,Oller:1997ng}. 
Thus, due to the equal masses of $f_0(980)$ and $a_0(980)$ close to the $K\bar{K}$ threshold, the $a_0(980)$-$f_0(980)$ mixing effect was predicted more than 40 years ago~\cite{Achasov:1979xc}. 
Later, more theoretical work discussing the $a_0(980)$-$f_0(980)$ mixing can be found in Refs.~\cite{Kerbikov:2000pu,Close:2000ah,Grishina:2001zj,Achasov:2002hg,Kudryavtsev:2002uu,Achasov:2003se,Wu:2007jh,Hanhart:2007bd,Wu:2008hx}, and the experimental evidence was reported by the BESIII Collaboration~\cite{Ablikim:2010aa,Ablikim:2018pik}. 
In fact, for equal masses and lacking an ideal decay channel, the observation of the $a_0(980)$-$f_0(980)$ mixing is very challenging. 
But, the case for $a_0(1710)$-$f_0(1710)$ mixing is different, which can potentially be observed in the experiments, and studying this mixing effect is the motivation of the present work. 

In the molecular picture, the interaction of $K^*\bar{K}^*$ is analogous to that of $K\bar{K}$, where one would expect that similar resonances emerge. 
The $f_0(1710)$ state was firstly observed about 40 years ago~\cite{Etkin:1981sg,Edwards:1981ex}. At first, it was interpreted as a normal light scalar meson in the conventional quark model~\cite{Godfrey:1985xj,Segovia:2008zza}, and an isovector partner state of 1.78 GeV was predicted. 
Furthermore, the $f_{0}(1710)$ was also regarded as a candidate of the scalar glueball~\cite{Close:2005vf,Giacosa:2005zt,Cheng:2006hu,Albaladejo:2008qa,Gui:2012gx,Fariborz:2015dou,Janowski:2014ppa,BESIII:2022riz}. 
But, as pointed out in Refs.~\cite{Chanowitz:2005du,Chao:2007sk}, the $f_{0}(1710)$ may contain large $s\bar{s}$ quarks components because it mainly decays to the $K\bar{K}$ and $\eta\eta$ channels. 
Thus, from the coupled channel interaction's point of view, it was considered as a molecular state of $K^*\bar{K}^*$ in Ref.~\cite{Geng:2008gx}, where an isovector partner $a_0$ state around $1.78$ GeV was predicted. 
Similar prediction from the $K^*\bar{K}^*$ interactions also can be found in Refs.~\cite{Du:2018gyn,Wang:2021jub}. 
The possibility, that $f_0(1710)$ being a molecular state was dynamically generated from the
vector-vector coupled channel interactions, was discussed in Refs.~\cite{Nagahiro:2008bn,Branz:2009cv,Geng:2010kma,Wang:2011tm,MartinezTorres:2012du,Xie:2014gla,Dai:2015cwa,Molina:2019wjj}.

Therefore, searching for $f_0(1710)$'s isovector partner is crucial to pin down whether it is a glueball or molecular state. 
Until 2021, a new state $a_{0}(1700)$ in the $\pi\eta$ invariant mass spectrum was observed in the $\eta_{c}\rightarrow\eta\pi^{+}\pi^{-}$ decay by the BABAR Collaboration~\cite{BaBar:2021fkz}, of which the mass and the width were given by $(1.704\pm0.005)\gev$ and $(0.110\pm0.019)\gev$, respectively. 
Moreover, the $f_{0}(1710)$ was also observed in the decays $\eta_{c} \to \eta^\prime K^+K^-(\pi^+\pi^-)$~\cite{BaBar:2021fkz}, which was consistent with the observation of the radiative decays $\Upsilon \to \gamma K^+K^- (\pi^+\pi^-)$~\cite{BaBar:2018uqa}. 
Furthermore, the decay $D_{s}^{+} \rightarrow \pi^{+}K_{S}^{0}K_{S}^{0}$ was investigated by the BESIII Collaboration~\cite{BESIII:2021anf}, where a resonance structure in the energy region around $1.710$ GeV was observed in the $K_{S}^{0}K_{S}^{0}$ mass distribution, with mass $M_{S(1710)} = (1.723\pm0.011)\gev$ and width $\Gamma_{S(1710)}= (0.140\pm0.015)\gev$. 
Note that, due to the $a_{0}(1710)$ and $f_{0}(1710)$ have identical quantum number $J^{PC}=0^{++}$ and their masses are assumed to be identical in Ref.~\cite{BESIII:2021anf}, thus they were not distinguished and assigned as S(1710) in this reference, which is in analogy to $S(980)$ for the states $a_{0}(980)$ and $f_{0}(980)$.
In fact the $S(1710)$ and $S(980)$ were the ``mixing" signals in the decay $D_{s}^{+} \rightarrow \pi^{+}K_{S}^{0}K_{S}^{0}$. 
Indeed, the admixture signal of $S(980)$ was also observed in the decay $D^+_s \rightarrow K^+ K^- \pi^+$~\cite{BaBar:2010wqe,BESIII:2020ctr}, where the $f_{0}(1710)$ was also seen. 
Due to the overlap of $a_{0}(1710)^0$ and $f_{0}(1710)$, the BESIII Collaboration continued to measure the related decay $D_{s}^{+} \rightarrow K_{S}^{0}K^{+}\pi^{0}$~\cite{BESIII:2022npc}, and observed a resonance $a_{0}(1710)^{+}$ in the $K_{S}^{0}K^{+}$ invariant mass spectrum, $M_{a_{0}(1710)} = (1.817\pm0.022)\gev$ and $\Gamma_{a_{0}(1710)}= (0.097\pm0.027)\gev$, which was called as the $a_{0}(1817)$ state in the existing literature
\footnote{In the updated online version of Particle Data Group (PDG)~\cite{Workman:2022ynf}, it was named as $a_{0}(1710)$.}. 

As suggested in Ref.~\cite{Guo:2022xqu}, the finding of Ref.~\cite{BESIII:2022npc} should be properly named as $a_{0}(1817)$, which was arranged in the same Regge trajectory with the $a_{0}(980)$, 
whereas, it was assigned as $a_{0}(1710)$ in Ref.~\cite{Wang:2022pin} from the coupled channel interaction of $K^*\bar{K}^*$. 
Using the coupled channel for the final state interaction, the decay $D_{s}^{+} \rightarrow K_{S}^{0}K^{+}\pi^{0}$ was investigated in Refs.~\cite{Zhu:2022guw,Wang:2023aza}, where the new state of Ref.~\cite{BESIII:2022npc} was assumed as the $a_{0}(1710)$. 
Note that, a combined fit for the invariant mass distributions was performed in Ref.~\cite{Wang:2023aza}, where a pole $(1.7936 + 0.0094 i)\gev$ was found for the $a_{0}(1710)$. 
Also applying the final state interaction approach, the decay $D_{s}^{+} \rightarrow \pi^{+}K_{S}^{0}K_{S}^{0}$ was studied in Refs.~\cite{Dai:2021owu,Zhu:2022wzk}, $\eta_c \to \bar{K}^0 K^+ \pi^-$ in Ref.~\cite{Ding:2023eps} and $J/\psi \to \bar{K}^0 K^+ \rho^-$ in Ref.~\cite{Ding:2024lqk}.  
With the MIT bag model, it was found in Ref.~\cite{Achasov:2023izs} that the strong $a_{0}$ coupling to the vector channels, such as $K^*\bar{K}^*$, $\rho \phi$, and so on, depending on the mass of the $a_{0}$, and thus, these vector channels should be detected to understand the nature of the new $a_{0}$. 
Furthermore, more comments on the $a_{0}(1710)$ can be found in Ref.~\cite{Oset:2023hyt} with some proposals for future experiments. 
Thus, in the present work, to understand more about the nature of $a_{0}(1710)$, we investigate the decay $D_{s}^{+} \rightarrow \pi^{+}K_{S}^{0}K_{S}^{0}$ with the final state interaction formalism. Note that, compared to the ones done in Refs.~\cite{Dai:2021owu,Zhu:2022wzk}, where the Breit-Wigner amplitude was used for the $a_{0}(1710)$ state. In this work, we use eleven coupled channels to dynamically reproduce the $a_{0}(1710)$ and also make a combined fit for the invariant mass distributions, where the mixing effect is found, see our results later.

The rest of the paper is organized as follows. 
In Sec.~\ref{Sec2}, we introduce the final state interaction formalism for the $D^+_s \rightarrow K^0_S K^0_S \pi^+$ decay in detail. 
The results of combined fits for the $K^0_S K^0_S$ and $K^0_S \pi^+ $ invariant mass distributions, and the branching ratios of different intermediate resonances are presented in Sec.~\ref{Sec3}. 
Then, a short summary is made in Sec.~\ref{Sec4}.

\section{Theoretical Framework}\label{Sec2}

In this work, we investigate the three-body weak decay $D_{s}^{+}\rightarrow K_{S}^{0}K_{S}^{0}\pi^{+}$. 
Firstly, we consider the dynamics at the quark level with the external and internal emission mechanisms of $W$-boson~\cite{Chau:1987tk,Chau:1982da,Morrison:1989xq}. 
Then, at the hadron level, we take into account the final state interactions in the $S$ wave, and the contribution from the vector meson resonance produced in the $P$ wave. 

\begin{figure}[!htbp]
\centering
\includegraphics[scale=0.4]{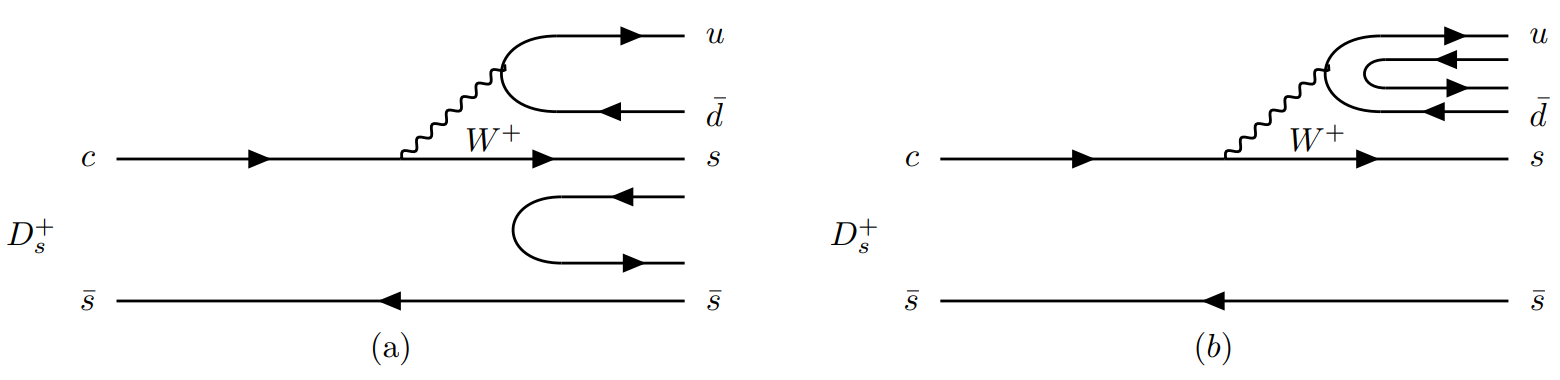}
\captionsetup{justification=raggedright}
\caption{$W$-external emission mechanism for the $D_{s}^{+}\rightarrow K_{S}^{0}K_{S}^{0}\pi^{+}$ decay. (a) The $s \bar{s}$ quark pair hadronizes into final mesons, (b) the $u \bar{d}$ quark pair hadronizes into final mesons.} \label{Feyman1}
\end{figure}

\begin{figure}[!htbp]
\centering
\includegraphics[scale=0.4]{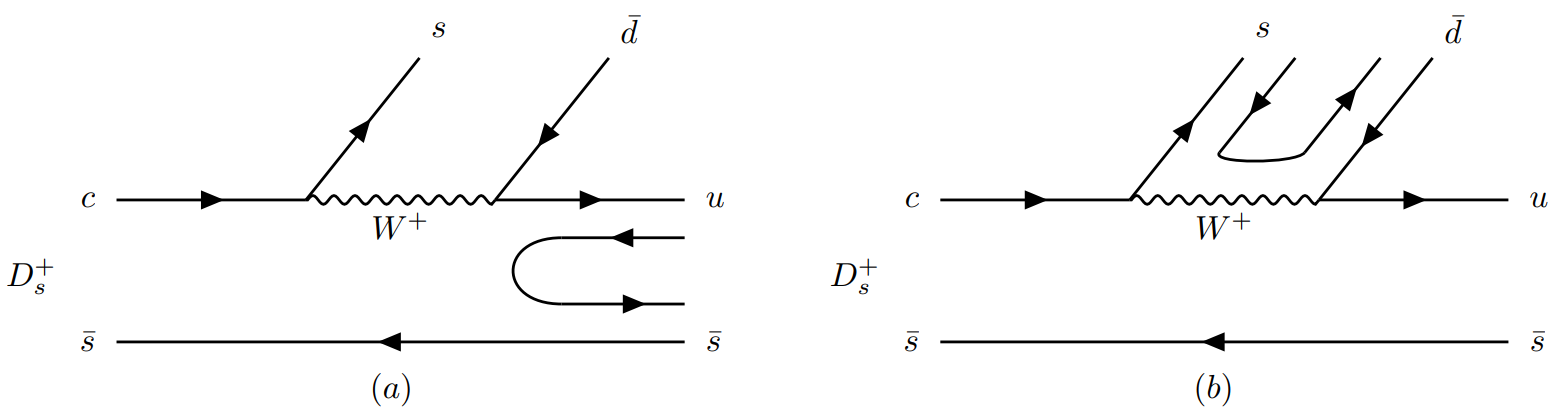}
\captionsetup{justification=raggedright}
\caption{$W$-external emission mechanism for the $D_{s}^{+}\rightarrow K_{S}^{0}K_{S}^{0}\pi^{+}$ decay. (a) The $u \bar{s}$ quark pair hadronizes into final mesons, (b) the $s \bar{d}$ quark pair hadronizes into final mesons.} \label{Feyman2}
\end{figure}

For the $D_{s}^{+}\rightarrow K_{S}^{0}K_{S}^{0}\pi^{+}$ decay, we only consider the contributions of the weak decay topology from the $W$-external and -internal emission mechanisms, which are the dominant ones. The related Feynman diagrams are shown in Figs.~\ref{Feyman1} and \ref{Feyman2}. 
In Fig.~\ref{Feyman1}, the $c$ quark coming from $D_{s}^{+}$ decays into $W^{+}$ boson and $s$ quark, and then the $W^{+}$ boson decays into a $u\bar{d}$ quark pair, while the $\bar{s}$ quark remains unchanged. 
There are two possible hadronization processes.
First, the $u\bar{d}$ quark pair forms a $\pi^{+}$ or $\rho^{+}$ meson directly, while the $s\bar{s}$ quark pair decays into two mesons via hadronization with $q\bar{q}=u\bar{u}+d\bar{d}+s\bar{s}$ produced from the vacuum. 
Second, the $s\bar{s}$ quark pair goes into a $\phi$ or $\eta$ meson, while the $u\bar{d}$ quark pair created by the $W^{+}$ boson undergoes the hadronization. 
The corresponding processes for these hadronizations can be written as,
\begin{equation}
\begin{aligned}
\left|H^{(1a)}\right\rangle = & V_{P}^{(1a)}V_{cs}V_{ud}(u\bar{d}\rightarrow\pi^{+})\left|s(\bar{u}u+\bar{d}d+\bar{s}s)\bar{s}\right\rangle  \\ 
&+ V_{P}^{*(1a)}V_{cs}V_{ud}(u\bar{d}\rightarrow\rho^{+})\left|s(\bar{u}u+\bar{d}d+\bar{s}s)\bar{s}\right\rangle  \\
=& V_{P}^{(1a)}V_{cs}V_{ud}(\pi^{+})(M\cdot M)_{33}+V_{P}^{*(1a)}V_{cs}V_{ud}(\rho^{+})(M\cdot M)_{33},
\end{aligned}
\label{eq1}
\end{equation}

\begin{equation}
\begin{aligned}
\left|H^{(1b)}\right\rangle =& V_{P}^{(1b)}V_{cs}V_{ud}(s\bar{s}\rightarrow-\frac{2}{\sqrt{6}}\eta)\left|u(\bar{u}u+\bar{d}d+\bar{s}s)\bar{d}\right\rangle  \\
&+ V_{P}^{*(1b)}V_{cs}V_{ud}(s\bar{s}\rightarrow\phi)\left|u(\bar{u}u+\bar{d}d+\bar{s}s)\bar{d}\right\rangle  \\
=& V_{P}^{(1b)}V_{cs}V_{ud}(-\frac{2}{\sqrt{6}}\eta)(M\cdot M)_{12}+V_{P}^{*(1b)}V_{cs}V_{ud}(\phi)(M\cdot M)_{12},
\end{aligned} 
\label{eq2}
\end{equation}
where the factors $V_{P}^{(1a)}$, $V_{P}^{*(1a)}$, $V_{P}^{(1b)}$ and $V_{P}^{*(1b)}$ are the weak interaction strengths of the production vertices to generate $\pi^{+},\rho^{+},\eta$ and $\phi$ mesons, respectively~\cite{Ahmed:2020qkv,Liang:2015qva}. 
These factors contain all dynamical information and can be regarded as the constants determined by the experimental data fits. 
The factor $-\frac{2}{\sqrt{6}}$ is due to the flavor component $s\bar{s}$ of $\eta$ meson, where we take $\eta = \eta_8$ 
\footnote{If one consider the $\eta$-$\eta^\prime$ mixing, the matrix $P$ of Eq.~\eqref{eq7} will be different, see Ref.~\cite{Wang:2022aga}.}. 
The factors $V_{cs}$ and $V_{ud}$ are the elements from the Cabibbo-Kobayashi-Maskawa (CKM) matrix. 
The elements of $q\bar{q}$ quark pairs can form a matrix $M$ in SU(3), which is defined as,
\begin{eqnarray}
M=\left( 
\begin{array}{ccc}
u\bar{u} & u\bar{d} & u\bar{s} \\ 
d\bar{u} & d\bar{d} & d\bar{s} \\ 
s\bar{u} & s\bar{d} & s\bar{s}%
\end{array}%
\right).
\label{eq5}
\end{eqnarray}%

\begin{figure}[!htbp]
\begin{subfigure}{0.55\textwidth}
\centering
\includegraphics[width=1\linewidth]{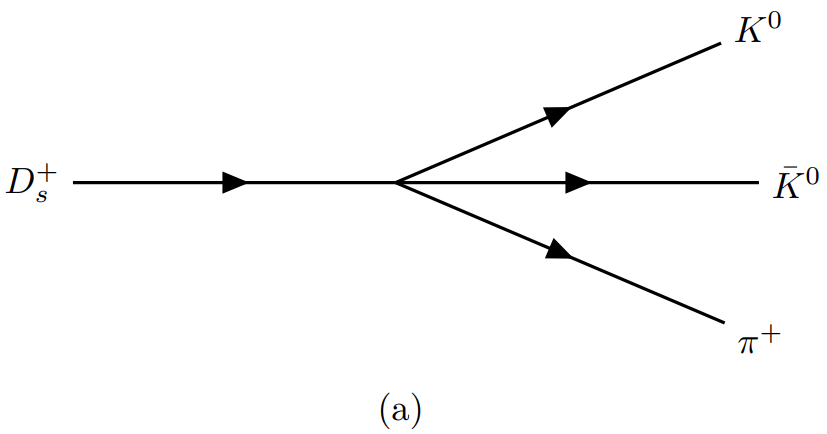} 
%\caption{Tree-level diagram.}
\label{Rescatter1}
\end{subfigure}
%\quad
\begin{subfigure}{0.475\textwidth}  
\centering 
\includegraphics[width=1\linewidth]{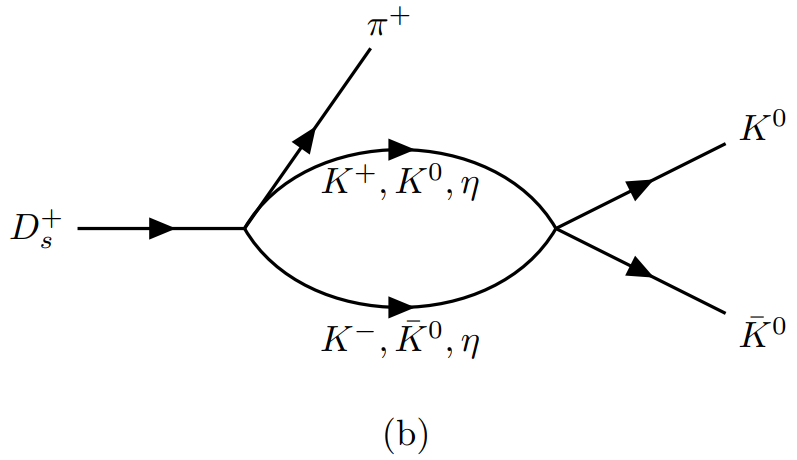} 
%\caption{Rescattering of $K^+ K^-$, $K^0 \bar{K}^0$ and $\eta \eta$.}
\label{Rescatter2} 
\end{subfigure}	
%\quad
\begin{subfigure}{0.475\textwidth}  
\centering
\includegraphics[width=1\linewidth]{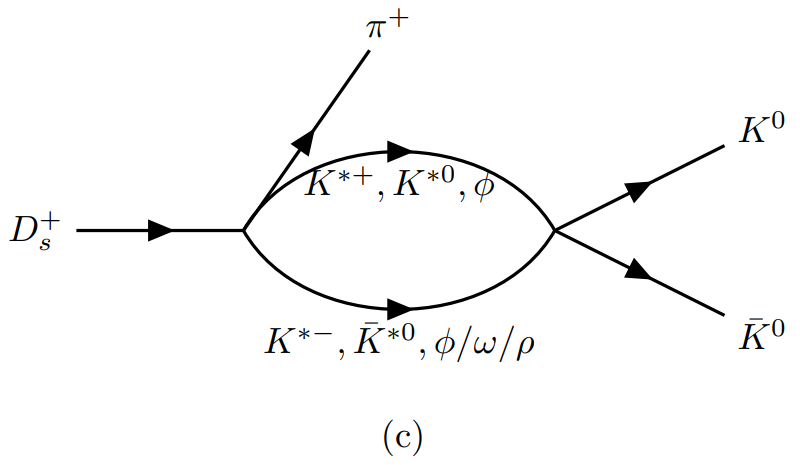} 
%\caption{Rescattering of $K^{*+} K^{*-}$, $K^{*0} \bar{K}^{*0}$ and $\phi \phi$, $\phi \omega$ and $\phi \rho$.}
\label{Rescatter3}
\end{subfigure}
\captionsetup{justification=raggedright}
\caption{Diagrammatic representations of the $D_{s}^{+}\rightarrow K_{S}^{0}K_{S}^{0}\pi^{+}$ decay. (a) Tree-level diagram. (b) Rescattering of $K^+ K^-$, $K^0 \bar{K}^0$ and $\eta \eta$. (c) Rescattering of $K^{*+} K^{*-}$, $K^{*0} \bar{K}^{*0}$ , $\phi \phi$, $\phi \omega$ and $\phi \rho$.}
\label{Rescatter}
\end{figure}

Similarly, the case of the $W$-internal emission is shown in Fig.~\ref{Feyman2}, which also contains two possible hadronization processes. 
First, the $s\bar{d}$ quark pair goes in to a $\bar{K}^{0}$ or  $\bar{K}^{*0}$ meson, the $u\bar{s}$ quark pair hadronizes with the $q\bar{q}$ quark pairs produced from the vacuum. 
Second, the $u\bar{s}$ quark pair forms a $K^{+}$ or  $K^{*+}$ meson, while the $s\bar{d}$ quark pair hadronizes into two final states with the $q\bar{q}$ quark pairs produced from the vacuum. 
Then, these processes can be written as,

\begin{equation}
\begin{aligned}
\left|H^{(2a)}\right\rangle = & V_{P}^{(2a)}V_{cs}V_{ud}(s\bar{d}\rightarrow\bar{K}^{0})\left|u(\bar{u}u+\bar{d}d+\bar{s}s)\bar{s}\right\rangle  \\
& +V_{P}^{*(2a)}V_{cs}V_{ud}(s\bar{d}\rightarrow\bar{K}^{*0})\left|u(\bar{u}u+\bar{d}d+\bar{s}s)\bar{s}\right\rangle  \\
=& V_{P}^{(2a)}V_{cs}V_{ud}(\bar{K}^{0})(M\cdot M)_{13}+V_{P}^{*(2a)}V_{cs}V_{ud}(\bar{K}^{*0})(M\cdot M)_{13},
\end{aligned} 
\label{eq3}
\end{equation}

\begin{equation}
\begin{aligned}
\left|H^{(2b)}\right\rangle =& V_{P}^{(2b)}V_{cs}V_{ud}(u\bar{s}\rightarrow K^{+} )\left|s(\bar{u}u+\bar{d}d+\bar{s}s)\bar{d}\right\rangle  \\
&+ V_{P}^{*(2b)}V_{cs}V_{ud}(u\bar{s}\rightarrow K^{*+} )\left|s(\bar{u}u+\bar{d}d+\bar{s}s)\bar{d}\right\rangle  \\
=& V_{P}^{(2b)}V_{cs}V_{ud}( K^{+} )(M\cdot M)_{32}+V_{P}^{*(2b)}V_{cs}V_{ud}( K^{*+} )(M\cdot M)_{32},
\end{aligned} 
\label{eq4}
\end{equation}
where the factors $V_{P}^{(2a)}$, $V_{P}^{*(2a)}$, $V_{P}^{(2b)}$, $V_{P}^{*(2b)}$ are the production vertices to generate $\bar{K}^{0}$, $\bar{K}^{*0}$, $K^{+}$ and $K^{*+}$ mesons, respectively. 
Subsequently, after the hadronization, the matrix $M$ at the quark level can be transferred to the hadron level form in the terms of the pseudoscalar ($P$) or vector ($V$) mesons, rewritten as, 

\begin{eqnarray}
P = \left( 
\begin{array}{ccc}
\frac{1}{\sqrt{2}}\pi ^{0}+\frac{1}{\sqrt{6}}\eta  & \pi ^{+} & K^{+} \\ 
\pi ^{-} & -\frac{1}{\sqrt{2}}\pi ^{0}+\frac{1}{\sqrt{6}}\eta  & K^{0} \\ 
K^{-} & \bar{K}^{0} & -\frac{2}{\sqrt{6}}\eta 
\end{array}%
\right), 
\label{eq7}
\end{eqnarray}%

\begin{eqnarray}
V=\left(\begin{array}{ccc}
\frac{1}{\sqrt{2}}\rho^{0}+\frac{1}{\sqrt{2}}\omega & \rho^{+} & K^{*+}\\
\rho^{-} & -\frac{1}{\sqrt{2}}\rho^{0}+\frac{1}{\sqrt{2}}\omega & K^{*0}\\
K^{*-} & \bar{K}^{*0} & \phi
\end{array}\right),
\label{eq6}
\end{eqnarray}%
where we take $\eta \equiv \eta_{8}$ as done in Ref.~\cite{Liang:2014tia,Toledo:2020zxj}. 
It should be mentioned that the component $(M\cdot M)_{ij}$ has four situations with two matrices of physical meson, such as $(P\cdot P)_{ij}$, $(V\cdot V)_{ij}$, $(V\cdot P)_{ij}$ and $(P\cdot V)_{ij}$. 
Then, one can rewrite the hadronization processes mentioned above, where the explicit formulas are given by
\begin{equation}
\begin{aligned}
\left|H^{(1a)}\right\rangle  = &V_{cs}V_{ud} V_{P}^{(1a)}\left(\pi^{+}K^{+}K^{-}+\pi^{+}K^{0}\bar{K}^{0}+\frac{2}{3}\eta\eta\pi^{+}\right)  \\
&+ V_{cs}V_{ud} V_{P}^{'(1a)} \left(\pi^{+}K^{*+}K^{*-}+\pi^{+}K^{*0}\bar{K}^{*0}+\phi\phi\pi^{+}\right) ,  \\
\end{aligned}
\label{eq8}
\end{equation}

\begin{equation}
\begin{aligned}
\left|H^{(1b)}\right\rangle =& V_{cs}V_{ud}V_{P}^{(1b)}\left(-\frac{2}{3}\eta\eta\pi^{+}\right)+V_{cs}V_{ud}\hat{V}_{P}^{*(1b)}\left(-\frac{1}{\sqrt{2}}\pi^{+}\rho^{0}\phi+\frac{1}{\sqrt{2}}\pi^{+}\omega\phi\right) \\
&+V_{cs}V_{ud}\bar{V}_{P}^{*(1b)}\left(\frac{1}{\sqrt{2}}\pi^{+}\rho^{0}\phi+\frac{1}{\sqrt{2}}\pi^{+}\omega\phi\right) ,
\end{aligned}
\label{eq9}
\end{equation}

\begin{equation}
\begin{aligned}
\left|H^{(2a)}\right\rangle =V_{cs}V_{ud}V_{P}^{(2a)}\left(\pi^{+}K^{0}\bar{K}^{0}\right)+V_{cs}V_{ud}\hat{V}_{P}^{*(2a)}\left(\pi^{+}K^{*0}\bar{K}^{*0}\right) ,
\end{aligned}
\label{eq10}
\end{equation}

\begin{equation}
\begin{aligned}
\left|H^{(2b)}\right\rangle =V_{cs}V_{ud}V_{P}^{(2b)}\left(\pi^{+}K^{+}K^{-}\right)+V_{cs}V_{ud}\bar{V}_{P}^{*(2b)}\left(\pi^{+}K^{*+}K^{*-}\right) ,
\end{aligned}
\label{eq11}
\end{equation}
where we only keep the terms that contribute to the final states $K_{S}^{0}K_{S}^{0}\pi^{+}$. 
Besides, in order to distinguish the contribution from the different terms of $(M \cdot M)_{ij}$, we adopt $V_{P}^{(1a/1b/2a/2b)}$, $V_{P}^{'(1a/1b/2a/2b)}$, $\bar{V}_{P}^{(1a/1b/2a/2b)}$ and $\hat{V}_{P}^{(1a/1b/2a/2b)}$ to represent the $(P\cdot P)_{ij}$, $(V\cdot V)_{ij}$, $(V\cdot P)_{ij}$ and $(P\cdot V)_{ij}$, respectively, where the superscripts $1a/1b/2a/2b$ denote the contributions of the corresponding subfigures of Figs.~\ref{Feyman1} and \ref{Feyman2}, respectively. 
Then we obtain the total contributions in the S-wave,

\begin{equation}
\begin{aligned}
\left|H\right\rangle = &\left|H^{(1a)}\right\rangle +\left|H^{(1b)}\right\rangle +\left|H^{(2a)}\right\rangle +\left|H^{(2b)}\right\rangle  \\
=& V_{cs}V_{ud}\left(V_{P}^{(1a)}+V_{P}^{(2b)}\right) \pi^{+}K^{+}K^{-}+V_{cs}V_{ud}\left(V_{P}^{(1a)}+V_{P}^{(2a)}\right) \pi^{+}K^{0}\bar{K}^{0}  \\
&+ \frac{2}{3}V_{cs}V_{ud}\left(V_{P}^{(1a)}-V_{P}^{(1b)}\right) \pi^{+}\eta\eta+V_{cs}V_{ud}\left(V_{P}^{'(1a)}+\bar{V}_{P}^{*(2b)}\right)\ \pi^{+}K^{*+}K^{*-}  \\
&+ V_{cs}V_{ud}\left(V_{P}^{'(1a)}+\hat{V}_{P}^{*(2a)}\right) \pi^{+}K^{*0}\bar{K^{*0}}+V_{cs}V_{ud} V_{P}^{'(1a)}  \pi^{+}\phi\phi  \\
&+ \frac{1}{\sqrt{2}}V_{cs}V_{ud}\left(\hat{V}_{P}^{*(1b)}+\bar{V}_{P}^{*(1b)}\right)\pi^{+}\omega\phi+\frac{1}{\sqrt{2}}V_{cs}V_{ud}\left(-\hat{V}_{P}^{*(1b)}+\bar{V}_{P}^{*(1b)}\right) \pi^{+}\rho^{0}\phi  \\
=& C_{1}\pi^{+}K^{+}K^{-} + C_{2} \pi^{+}K^{0}\bar{K^{0}} + \frac{2}{3} C_{3} \pi^{+}\eta\eta + C_{4} \pi^{+}K^{*+}K^{*-} + C_{5} \pi^{+}K^{*0}\bar{K}^{*0} \\
&+ C_{6} \pi^{+}\phi\phi + \frac{1}{\sqrt{2}}C_{7} \pi^{+}\omega\phi + \frac{1}{\sqrt{2}} C_{8} \pi^{+}\rho^{0}\phi,
\label{eq12}
\end{aligned}
\end{equation}
where the factors $C_1$, $C_2$, $C_3$, $C_4$, $C_5$, $C_6$, $C_7$, and $C_8$ are defined as $C_{1} = V_{cs}V_{ud}\left(V_{P}^{(1a)}+V_{P}^{(2b)}\right)$, $C_{2} = V_{cs}V_{ud} \left(V_{P}^{(1a)}+V_{P}^{(2a)}\right)$, $C_{3} = V_{cs}V_{ud} \left(V_{P}^{(1a)}-V_{P}^{(1b)}\right)$, $C_{4} = V_{cs}V_{ud} \left(V_{P}^{'(1a)}+\bar{V}_{P}^{*(2b)}\right)$, $C_{5} = V_{cs}V_{ud} \left(V_{P}^{'(1a)}+\hat{V}_{P}^{*(2a)}\right)$, $C_{6}= V_{cs}V_{ud}\left(V_{P}^{'(1a)}\right)$, $C_{7} = V_{cs}V_{ud}\left(\hat{V}_{P}^{*(1b)}+\bar{V}_{P}^{*(1b)}\right)$ and $C_{8} = V_{cs}V_{ud}\left(-\hat{V}_{P}^{*(1b)}+\bar{V}_{P}^{*(1b)}\right)$. 
Note that the elements of the CKM matrix and the vertex factors of weak decay are absorbed in these factors $C_i$.  
In our formalism, these parameters are taken as free constants, which also include a global normalization factor to match the events of the experimental data when we determine them by fitting the invariant mass distributions later. 
From Eq.~(\ref{eq12}), one can see that the final states $K_{S}^{0}K_{S}^{0}\pi^{+}$ can produce directly in the hadronization processes, while the remaining part can get these final states via rescattering procedures, which are shown in Fig.~\ref{Rescatter}. 

Therefore, under the dominant $W$-external and -internal emission mechanisms, the amplitude of $D_{s}^{+}\rightarrow K_{S}^{0}K_{S}^{0}\pi^{+}$ decay process in the $S$-wave can be written as,
\begin{equation}
\begin{aligned}
t(M_{12}) \vert_{K^0 \bar{K}^0 \pi^+} = & C_{1}G_{K^{+}K^{-}}(M_{12}) T_{K^{+}K^{-}\rightarrow K^{0}\bar{K}^{0}}(M_{12}) + C_{2}+C_{2}G_{K^{0}\bar{K}^{0}}(M_{12}) T_{K^{0}\bar{K}^{0}\rightarrow K^{0}\bar{K}^{0}}(M_{12}) \\
&+ \frac{2}{3}C_{3}G_{\eta\eta}(M_{12}) T_{\eta\eta\rightarrow K^{0}\bar{K}^{0}}(M_{12}) + C_{4}G_{K^{*+}K^{*-}}(M_{12}) T_{K^{*+}K^{*-}\rightarrow K^{0}\bar{K}^{0}}(M_{12})  \\
&+ C_{5}G_{K^{*0}\bar{K}^{*0}}(M_{12})T_{K^{*0}\bar{K}^{*0}\rightarrow K^{0}\bar{K}^{0}}(M_{12}) + C_{6}G_{\phi\phi}(M_{12}) T_{\phi\phi\rightarrow K^{0}\bar{K}^{0}}(M_{12})  \\
&+\frac{1}{\sqrt{2}}C_{7}G_{\omega\phi}(M_{12})T_{\omega\phi\rightarrow K^{0}\bar{K}^{0}}(M_{12}) + \frac{1}{\sqrt{2}}C_{8}G_{\rho^{0}\phi}(M_{12})T_{\rho^{0}\phi\rightarrow K^{0}\bar{K}^{0}}(M_{12}),
\label{eq13}
\end{aligned}
\end{equation}
where $M_{ij}$ is the energy of two particles in the center-of-mass (CM) frame, and the subscripts $i, j = 1, 2, 3$ denote three final states of $K^0$, $\bar{K}^0$ and $\pi^+$, respectively. 
Then, we take $\left|K_{S}^{0}\right\rangle =\frac{1}{\sqrt{2}}\left(\left|K^{0}\right\rangle -\left|\bar{K}^{0}\right\rangle \right)$ into account~\cite{Dai:2021owu} and change the final states from $K^{0}$ and $\bar{K}^{0}$ to $K_{S}^{0}$, and thus, the Eq.~(\ref{eq13}) can be revised into 
\begin{equation}
\begin{aligned}
& t(M_{12}) \vert_{K_S^0 K_S^0 \pi^+} \\
&=  -\frac{1}{2}C_{1}G_{K^{+}K^{-}}(M_{12}) T_{K^{+}K^{-}\rightarrow K^{0}\bar{K}^{0}}(M_{12}) -\frac{1}{2}C_{2} -\frac{1}{2}C_{2}G_{K^{0}\bar{K}^{0}}(M_{12}) T_{K^{0}\bar{K}^{0}\rightarrow K^{0}\bar{K}^{0}}(M_{12}) \\
& \quad - \frac{1}{3}C_{3}G_{\eta\eta}(M_{12})T_{\eta\eta\rightarrow K^{0}\bar{K}^{0}}(M_{12}) - \frac{1}{2}C_{4}G_{K^{*+}K^{*-}}(M_{12})T_{K^{*+}K^{*-}\rightarrow K^{0}\bar{K}^{0}}(M_{12}) \\
& \quad -\frac{1}{2}C_{5}G_{K^{*0}\bar{K}^{*0}}(M_{12})T_{K^{*0}\bar{K}^{*0}\rightarrow K^{0}\bar{K}^{0}}(M_{12}) -\frac{1}{2}C_{6}G_{\phi\phi}(M_{12})T_{\phi\phi\rightarrow K^{0}\bar{K}^{0}}(M_{12}) \\
& \quad - \frac{1}{2\sqrt{2}}C_{7}G_{\omega\phi}(M_{12})T_{\omega\phi\rightarrow K^{0}\bar{K}^{0}}(M_{12}) -\frac{1}{2\sqrt{2}}C_{8}G_{\rho^{0}\phi}(M_{12})T_{\rho^{0}\phi\rightarrow K^{0}\bar{K}^{0}}(M_{12}),
\label{eq14}
\end{aligned}
\end{equation}
where $G_{PP^{'}(VV^{'})}$ and $T_{PP^{'}(VV^{'}) \rightarrow PP^{'}}$ are the loop functions and the two-body scattering amplitudes, respectively, see the detail below. 

The matrix $G$ is diagonal, which is made up of the meson-meson loop function. The explicit form of the loop function with the dimensional regularization is given by~\cite{Oller:2000fj,Oller:1998zr,Gamermann:2006nm,Alvarez-Ruso:2010rqm,Guo:2016zep},
\begin{equation}
\begin{aligned} 
G_{PP^{'}}\left( M_{inv}\right)  =&\frac{1}{16\pi ^{2}} \Big\{a_{\mu}\left( \mu \right)
+\ln \frac{m_{1}^{2}}{\mu ^{2}}+\frac{m_{2}^{2}-m_{1}^{2}+M^2_{inv}}{2M^2_{inv}}\ln \frac{
m_{2}^{2}}{m_{1}^{2}} \\ 
&+\frac{q_{cm}\left( M_{inv}\right) }{M_{inv}} \big[\ln \left( M^2_{inv}-\left(
m_{2}^{2}-m_{1}^{2}\right) +2q_{cm}\left( M_{inv}\right) M_{inv}\right)  \\ 
&+\ln \left( M^2_{inv}+\left( m_{2}^{2}-m_{1}^{2}\right) +2q_{cm}\left( M_{inv}\right) 
M_{inv}\right)  \\ 
&-\ln \left( -M^2_{inv}-\left( m_{2}^{2}-m_{1}^{2}\right) +2q_{cm}\left( M_{inv}\right) 
M_{inv}\right)  \\ 
&-\ln \left( -M^2_{inv}+\left( m_{2}^{2}-m_{1}^{2}\right) +2q_{cm}\left( M_{inv}\right) 
M_{inv}\right) \big]\Big\}, 
\end{aligned} \label{eq15}
\end{equation}
where $m_{1}$ and $m_{2}$ are the masses of two intermediate particles in the loop, $M_{inv}$ is the invariant mass of two mesons in the system, and $\mu$ is the regularization scale, of which the value will be determined by fitting experimental data in Sec.~\ref{Sec3}. 
Besides, $a(\mu)$ is the subtraction constant, of which the value can be evaluated by~\cite{Oller:2000fj,Duan:2020vye,Wang:2021kka}
\begin{equation}
a(\mu)=-2\ln\left(1+\sqrt{1+\frac{m_{1}^{2}}{\mu^{2}}}\right)+\cdots,
\label{eq16}
\end{equation}
where $m_{1}$ is the mass of a larger-mass meson in the corresponding channels, while the ellipses indicates the ignored higher order terms in the nonrelativistic expansion~\cite{Guo:2018tjx}. 
Furthermore, $q_{cm}\left(M_{inv}\right)$ is the three-momentum of the particles in the CM frame, 
\begin{equation}
q_{cm}\left(M_{inv}\right)=\frac{\lambda^{1/2}\left(M_{inv}^{2},m_{1}^{2},m_{2}^{2}\right)}{2M_{inv}} ,
\end{equation}
with the usual K\"allen triangle function $\lambda \left(a,b,c\right) = a^{2}+b^{2}+c^{2}-2\left(ab+ac+bc\right)$. 

\begin{figure}[!htbp]
\centering
\includegraphics[scale=0.35]{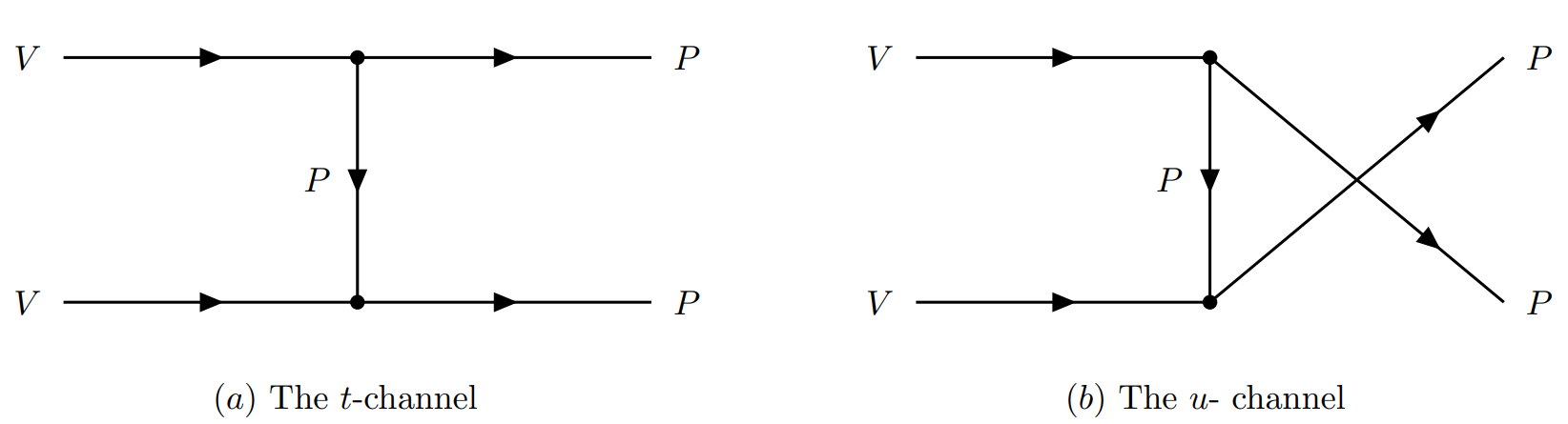}
\captionsetup{justification=raggedright}
\caption{Feynman diagrams of $t$- and $u$-channels.} \label{interaction}
\end{figure}

Moreover, the two-body scattering amplitude in Eq.~(\ref{eq14}) can be calculated by the coupled channel Bethe-Salpeter equation~\cite{Oller:1997ti,Oset:1997it},
\begin{equation}
T=\left[1-VG\right]^{-1}V ,
\end{equation}
where the matrix $G$ is constructed by the loop functions, the element of which is given by Eq.~(\ref{eq15}), and the matrix $V$ is made of the interaction potential for each coupled channel. 
For isospin $I=1$ sector, we consider six channels, $(1)\,K^{*}\bar{K}^{*}$, $(2)\,\rho\omega$, $(3)\,\rho \phi$, $(4)\,\rho \rho$, $(5)\,K\bar{K}$, $(6)\,\pi \eta$.
For isospin $I=0$ sector, we consider eight channels,$(1)\,K^{*}\bar{K}^{*}$, $(2)\,\rho\rho$, $(3)\,\omega\omega$, $(4)\,\omega\phi$, $(5)\,\phi\phi$, $(6)\,\pi\pi$, $(7)\,K\bar{K}$, $(8)\,\eta\eta$. 
Among them, the potentials $V_{VV \rightarrow VV}$ are taken from Ref.~\cite{Geng:2008gx}, which considered the tree-level transition of four-vector-contact diagrams and the $t/u$-channel vector meson-exchange terms. And the potentials $V_{PP \rightarrow PP}$ are taken from Refs.~\cite{Oller:1997ti,Duan:2020vye,Ling:2021qzl,Liang:2014tia,Wang:2021ews}, which only include the tree level contributions. 
Besides, the potentials $V_{VV \rightarrow PP}$ are calculated as done in Ref.~\cite{Wang:2022pin}, where the Feynman diagrams of $t$- and $u$-channels are considered, as shown in Fig.~\ref{interaction}. The Lagrangian for the $VPP$ vertex is given by~\cite{Bando:1984ej,Bando:1987br}  
\begin{equation}
\mathcal{L}_{VPP}=-ig\left\langle V_{\mu}\left[P,\partial^{\mu}P\right]\right\rangle, 
\end{equation}
where the symbol $\left\langle \ldots\right\rangle $ stands for the trace in color space, $g=M_V/(2f_{\pi})$ with $M_V=0.84566$ GeV, which is the averaged vector-meson mass, and $f_{\pi}=0.093$ GeV is pion decay constant, which is taken from Ref.~\cite{Wang:2022pin,Oller:1997ti}. 
Then, some parts of the potentials of $V_{VV \rightarrow PP}$ are given by (for simplify not listing them all),
\begin{equation}
\begin{aligned}
& V_{K^{*+}K^{*-}\rightarrow K^{0}\bar{K}^{0}} = -\frac{4}{t-m_{\pi}^{2}}g^{2}\epsilon _{1\mu }k_{3}^{\mu }\epsilon _{2\nu }k_{4}^{\nu },  \\
& V_{K^{*0}\bar{K}^{*0}\rightarrow K^{0}\bar{K}^{0}} = -2\left( \frac{%
3}{t-m_{\eta }^{2}}+\frac{1}{t-m_{\pi }^{2}}\right) g^{2}\epsilon _{1\mu
}k_{3}^{\mu }\epsilon _{2\nu }k_{4}^{\nu },  \\
& V_{\phi \phi \rightarrow K^{0}\bar{K}^{0}} = -4g^{2}\left( \frac{1}{%
t-m_{K}^{2}}\epsilon _{1\mu }k_{3}^{\mu }\epsilon _{2\nu }k_{4}^{\nu }+\frac{%
1}{u-m_{K}^{2}}\epsilon _{1\mu }k_{4}^{\mu }\epsilon _{2\nu }k_{3}^{\nu
}\right),  \\
& V_{\omega \phi \rightarrow K^{0}\bar{K}^{0}} = 2\sqrt{2}g^{2}\left( \frac{1%
}{t-m_{K}^{2}}\epsilon _{1\mu }k_{3}^{\mu }\epsilon _{2\nu }k_{4}^{\nu }+%
\frac{1}{u-m_{K}^{2}}\epsilon _{1\mu }k_{4}^{\mu }\epsilon _{2\nu
}k_{3}^{\nu }\right),  \\
& V_{\rho \phi \rightarrow K^{0}\bar{K}^{0}} = -2\sqrt{2}\left( \frac{1}{%
t-m_{K}^{2}}g^{2}\epsilon _{1\mu }k_{3}^{\mu }\epsilon _{2\nu }k_{4}^{\nu }+%
\frac{1}{u-m_{K}^{2}}g^{2}\epsilon _{1\mu }k_{4}^{\mu }\epsilon _{2\nu
}k_{3}^{\nu }\right), 
\end{aligned}
\end{equation}
where $t=(k_1 - k_3)^2$ and $u = (k_1 - k_4)^2$, $\epsilon_i$ is the polarization vector, and $k_i$ is the four-momentum of the corresponding particles, of which the subscript $i(i=1, 2, 3, 4)$ denotes the particles in scattering process $V(1)V(2) \rightarrow P(3)P(4)$. 
It is worth mentioning that, as done in Refs.~\cite{Wang:2021jub,Molina:2008jw}, we also introduce a monopole form factor for each $VPP$ vertex of the exchanged pseudoscalar meson, of which the explicit expression is given by,
\begin{equation}%
F=\frac{\varLambda^{2}-m_{ex}^{2}}{\varLambda^{2}-q^{2}} ,
\end{equation}
where $m_{ex}$ is the mass of the exchanged pseudoscalar meson, $q$ is the transferred momentum, and the value of parameter $\Lambda$ is empirically chosen as 1.0 GeV. 
After performing the partial wave projection, we can obtain the $S$-wave potentials $V_{VV \rightarrow PP}$.  
Note that these non-diagonal potentials from the transitions $V_{VV \rightarrow PP}$ are in fact weak compared to the diagonal ones from the vector meson exchanges.

\begin{figure}[!htbp]
\centering
\includegraphics[scale=0.45]{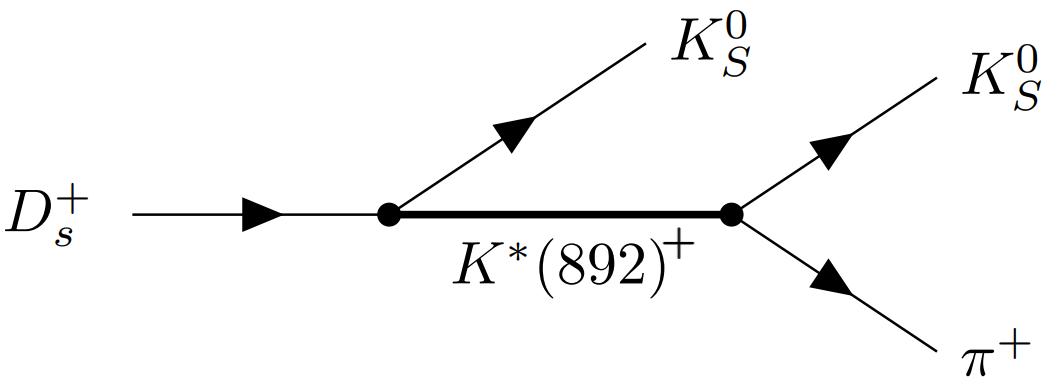}
\captionsetup{justification=raggedright}
\caption{Mechanism of $D^+_s \rightarrow K^0_S K^0_S \pi^+$ decay via the intermediate state $K^*(892)^+$.} \label{intermediate}
\end{figure}

Furthermore, we also consider the contribution of the vector resonance generated in the $P$ wave. 
As discussed above, the vector meson both with the $W$-external and -internal emission mechanisms can be produced directly. 
Thus, the vector meson coming from hadronization, such as $K^{*}(892)^{+}$, which is not generated in the meson-meson rescattering process, would also decays to the final states that we are looking for. 
The production mechanism is depicted in Fig.~\ref{intermediate}, and the relativistic amplitude for the decay $D_{s}^{+}\rightarrow K_{s}^{0}K^{*}(892)^{+}\rightarrow K_{S}^{0}K_{S}^{0}\pi^{+}$ can be written as~\cite{Toledo:2020zxj,Roca:2020lyi}
\begin{equation}
\begin{aligned}
t_{K^{*}(892)^{+}}(M_{12},M_{23}) =& \frac{\mathcal{D} e^{i\alpha_{K^{*}(892)^{+}}}}{M_{23}^{2}-M_{K^{*}(892)^{+}}^{2}+iM_{K^{*}(892)^{+}}\varGamma_{K^{*}(892)^{+}}} \\
& \times\left[\frac{(m_{D_{s}^{+}}^{2}-m_{K_{s}^{0}}^{2})(m_{K_{s}^{0}}^{2}-m_{\pi^{+}}^{2})}{M_{K^{*}(892)^{+}}^{2}}-M_{12}^{2}+M_{13}^{2}\right],
\label{eq22}
\end{aligned}
\end{equation}
where $\mathcal{D} $ and $ \alpha_{K^{*}(892)^{+}}$ are the normalization constant and the phase, respectively, which can be determined by fitting the experimental data. 
The mass and width of intermediate $K^*(892)^+$ are taken from the PDG~\cite{Workman:2022ynf}, of which the value are  $M_{K^*(892)^+} = 0.89167$ GeV and $\Gamma_{K^*(892)^+} = 0.0514$ GeV. 
Although there are three $s_{ij}$ variables, only two of them are independent, which fulfill the constraint condition,
\begin{equation}
M_{12}^{2}+M_{13}^{2}+M_{23}^{2}=m_{D_{s}^{+}}^{2}+m_{K_{S}^{0}}^{2}+m_{K_{S}^{0}}^{2}+m_{\pi^+}^{2}.
\end{equation}

Finally, the double differential width distribution for the three-body decay $D^+_s \rightarrow K^0_S K^0_S \pi^+$ can be calculated by~\cite{Workman:2022ynf},
\begin{equation}
\frac{d^{2}\Gamma}{dM_{12}dM_{23}}=\frac{1}{(2\pi)^{3}}\frac{M_{12}M_{23}}{8m_{D_{s}^{+}}^{3}} \frac{1}{2}  \mid \mathcal{M} \mid^{2},
\label{eq24}
\end{equation}
where the factor $1/2$ comes from the identical particle $K^0_S$ in the final states, and $\mathcal{M}$ is the total amplitude of the decay $D^+_s \rightarrow K^0_S K^0_S \pi^+$. From the discussions above, the total amplitude of $\mathcal{M}$ is written as
\begin{equation}
\mathcal{M} = t(M_{12}) \vert_{K_S^0 K_S^0 \pi^+} + t_{K^{*}(892)^{+}}(M_{12},M_{23}) + (1 \leftrightarrow 2),
\end{equation}
where the amplitude $t(M_{12}) \vert_{K_S^0 K_S^0 \pi^+}$ is given by Eq.~(\ref{eq14}), the one $t_{K^{*}(892)^{+}}(M_{12},M_{23})$ by Eq.~(\ref{eq22}), and $(1 \leftrightarrow 2)$ resembles the symmetry between the two identical  $K^0_S$ in the final states.
With Eq~(\ref{eq24}), one can easily obtain the invariant mass spectra $d\Gamma / dM_{12}$, $d\Gamma/dM_{13}$ and $d\Gamma/dM_{23}$ by integrating over each of the invariant mass variables with the limits of the Dalitz Plot, more details can be found in the PDG~\cite{Workman:2022ynf}.

\section{Results}\label{Sec3}

\begin{figure}[!htbp]
\begin{subfigure}{0.475\textwidth}
\centering
\includegraphics[width=1\linewidth]{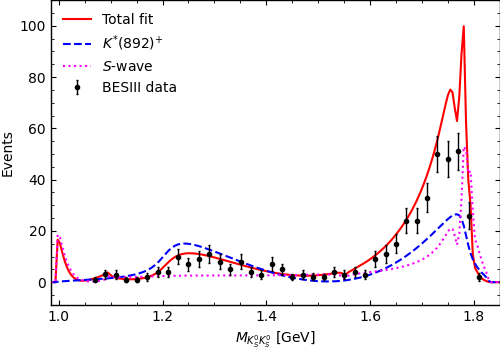} 
\caption{Invariant mass distribution of $K^0_S K^0_S$.}
\label{contribution1}
\end{subfigure}
%\quad
\begin{subfigure}{0.475\textwidth}  
\centering 
\includegraphics[width=1\linewidth]{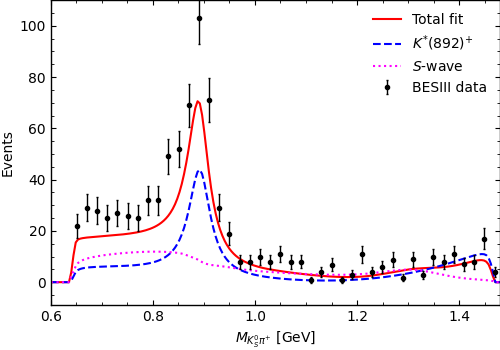} 
\caption{Invariant mass distribution of $K^0_S \pi^+$.}
\label{contribution2}  
\end{subfigure}	
\captionsetup{justification=raggedright}
\caption{Combined fit for the invariant mass distributions of the decay $D^+_s \rightarrow K^0_S K^0_S \pi^+$. The solid (red) line is the total contributions of the $S$ and $P$ waves, the dashed (blue) line represents the contribution of the $K^*(892)^+$, the dotted (magenta) lines is the contributions from the $S$-wave interactions with the resonances $S(980)$ and $S(1710)$. The dot (black) points are the experimental data measured by the BESIII Collaboration~\cite{BESIII:2021anf}.}
\label{Contribution}
\end{figure}

For the regularization scale $\mu$ in the loop functions, see Eq.~\eqref{eq15}, it is normal to set its value as $\mu=0.6$ GeV for the pseudoscalar-pseudoscalar interaction~\cite{Duan:2020vye,Wang:2021ews,Oller:1997ti,Dias:2016gou,Liang:2014tia} and $\mu=1.0$ GeV for the vector-vector meson interaction~\cite{Geng:2008gx}, respectively. 
One should keep in mind that $a_\mu$ and $\mu$ are not independent, see more discussions in Ref.~\cite{Ozpineci:2013zas}. 
Thus, in our formalism, due to including the interactions between the pseudoscalar and vector mesons, we take $\mu$ as a free parameter. 
Thus, in our theoretical formalism, there are eleven parameters: $\mu$ is the regularization scale in the loop functions, $C_i$ ($i=1,\ 2,\ \cdots,\ 8$) are the free parameters in $S$-wave interaction amplitude, $D$ and $\alpha_{K^*(892)^+}$ are the production factor and phase appearing in the $P$-wave amplitude. 
We can obtain these parameters by doing a combined fit to the invariant mass distributions of the $D^+_s \rightarrow K^0_S K^0_S \pi^+$ decay reported by the BESIII Collaboration~\cite{BESIII:2021anf}. 
The values of these parameters are given in Table~\ref{parameters}, where the fitted $\chi^{2}/dof.=175.97/(80-11)=2.55$. 
With these fitted parameters, the obtained results of the $K^0_S K^0_S$ and $K^0_S \pi^+$ invariant mass distributions are shown in Fig.~\ref{Contribution}. 
For the regularization scale, a value $\mu=0.648$ GeV is gotten from the fit, see Table~\ref{parameters}, and then, the subtraction constants $a_{\mu}$ for the corresponding coupled channels can be calculated by Eq.~(\ref{eq16}), obtained as
\begin{eqnarray}
\begin{aligned}
& a_{K^{*+}K^{*-}}=-1.99,\quad a_{K^{*0}\bar{K}^{*0}}=-1.99,\quad a_{\rho^0 \omega}=-1.89, \quad a_{\rho^0 \phi}=-2.10, \quad  a_{K^{+}K^{-}}=-1.63, \\ 
& a_{K^{0}\bar{K^{0}}}=-1.63,\quad a_{\pi^0 \eta}=-1.67, \quad a_{\rho^+ \rho^-}=-1.88, \quad a_{\rho^0 \rho^0}=-1.88, \quad a_{\omega \omega}=-1.89, \\
& a_{\omega \phi}=-2.10, \quad a_{\phi \phi}=-2.10, \quad a_{\pi^+ \pi^-}=-1.41, \quad a_{\pi^0 \pi^0}=-1.41, \quad a_{\eta \eta}=-1.67.
\end{aligned}
\label{eq:amu}
\end{eqnarray}

\begin{table}[!htb]
\centering
\caption{Values of the parameters from the fit.} \label{parameters}
\resizebox{0.75\textwidth}{!}
{\begin{tabular}{cccccccc}
%\begin{tabular*}{17.8cm}{@{\extracolsep{\fill}}*{7}{p{1.18cm}<{\centering}}}
\hline\hline
Parameters  & $\mu$     & $C_1$  & $C_2$ &  $C_3$   & $C_4$ &   $C_5$ & \\
Fit            & 0.648 GeV     & 8640.90 & 2980.71 & -1902.86 &  56906.35 &   -13433.15   \\
\hline
Parameters  & $C_6$ & $C_7$ & $C8$ &  $D$ & $\alpha_{K^*(892)^+}$  & $\chi^2/dof.$\\
Fit         &  -58284.22   & 102835.76  & 202807.71 & 54.80 & 0.0024 & 2.55 \\
\hline\hline
\end{tabular}}
\end{table}

\begin{figure}[htbp]
\begin{subfigure}{0.475\textwidth}
\centering
\includegraphics[width=1\linewidth]{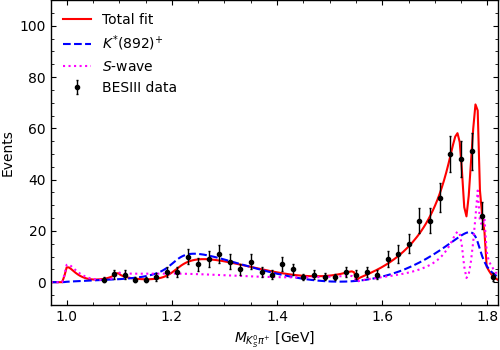} 
\caption{Invariant mass distribution of $K^0_S K^0_S$.}
\label{contribution1a}
\end{subfigure}
%\quad
\begin{subfigure}{0.475\textwidth}  
\centering 
\includegraphics[width=1\linewidth]{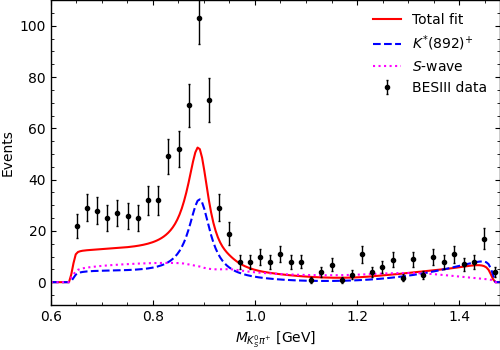} 
\caption{Invariant mass distribution of $K^0_S \pi^+$.}
\label{contribution2a}  
\end{subfigure}	
\captionsetup{justification=raggedright}
\caption{Fit only for the data of $K^0_S K^0_S$ spectrum. Others are the same as Fig.~\ref{Contribution}.}
\label{Contribution2x}
\end{figure} 

For the $K^0_S K^0_S$ invariant mass distributions as shown in Fig.~\ref{contribution1}, the bumps around 1.3 GeV benefit from the contribution of the $K^*(892)^+$ resonance, see the results of the dashed (blue) line, which also contributes to the region of $1.6$ - $1.8$ GeV. 
In Fig.~\ref{contribution1}, our fitting for the obvious resonance structure around $1.7$ GeV are a bit higher than the data, see the results of the solid (red) line, which also splits into two overlapped peaks due to the mixing effect of the $a_0(1710)$ and $f_0(1710)$ generated from the $S$-wave amplitude, and where it is difficult to fit well from our different tests, even though there are eight parameters in the $S$-wave amplitude, also see the discussions later.  
Furthermore, an obvious enhancement around the $K^0_S K^0_S$ threshold is caused by the resonances $S(980)$, which are also dynamically generated in the $S$-wave final state interactions. 
Then, using the obtained results of the $a_{\mu}$ for each couple channel, see Eq.~\eqref{eq:amu}, we calculate the corresponding poles for the $a_0(980)$, $f_0(980)$, $a_0(1710)$ and $f_0(1710)$ states in the complex Riemann sheets, which are listed in Table.~\ref{poles}. 
Since the pole is located at $(M + i \frac{\Gamma}{2})$, one can easily obtain the total width of the corresponding resonances. 
From Table~\ref{poles}, the pole for the $a_0(980)$, $f_0(980)$, and $a_0(1710)$ states are not much different from our former results of Ref.~\cite{Wang:2023aza,Ahmed:2020kmp}. 
However, the corresponding width of the $a_0(1710)$ is several times smaller than those obtained in Refs.~\cite{Geng:2008gx,Wang:2022pin,Du:2018gyn}. 
Note that there is a difference of about 30 MeV between the poles of $a_0(1710)$ and $f_0(1710)$, which with the small widths causes the split structure at the high energy region around $1.7\gev$ in the $K^0_S K^0_S$ mass distributions as discussed above. 
Similarly, for the $K^0_S \pi^+$ mass distribution as shown in Fig.~\ref{contribution2}, the contribution from the state $K^*(892)^+$ enhances in the energy region around 0.8 and $1.0\gev$, while the $S$-wave amplitude mainly contributes to the low-energy region. 
In Fig.~\ref{contribution2}, one can see that there are obvious discrepancies between our fit and experimental data. 
After checking carefully, we find that the data of the bumps around 1.3 GeV in Fig.~\ref{contribution1}, which suppress the $P$-wave amplitude of the $K^*(892)^+$ with the destroyed interference effect from the $S$-wave amplitude, also prevent the fitting results of Fig.~\ref{contribution2}. 
To reveal this suppression effect, we fit only the data of the invariant mass distribution of $K_{S}^{0}K_{S}^{0}$, as shown the results in Fig.~\ref{Contribution2x}.
As seen from Fig.~\ref{Contribution2x}, there are two obvious peaks in the energy region of $1.7\gev$ in Fig.~\ref{contribution1a}, which come from the contribution of the $S$-wave amplitude and show more clear mixing effect, and the data of the $K^0_S \pi^+$ spectrum in Fig.~\ref{contribution2a} is also described not bad and similar to what we get in Fig.~\ref{contribution2}. 
Note that using the fitting results of Fig.~\ref{Contribution2x}, the extracted poles and branching ratios are not much different from the results of Table~\ref{poles} and Eq.~\eqref{eq:bran} later.

\begin{table}[!htb]
\centering
\caption{Poles compared with the other works (unit: GeV).} \label{poles}
\resizebox{1\textwidth}{!}
{\begin{tabular}{cccccccc}
%\begin{tabular*}{17.8cm}{@{\extracolsep{\fill}}*{7}{p{1.18cm}<{\centering}}}
\hline\hline
 			   & This work          &  Ref.~\cite{Wang:2023aza}         &  Ref.~\cite{Ahmed:2020kmp}  &   Ref.~\cite{Geng:2008gx} &  Ref.~\cite{Wang:2022pin} &   Ref.~\cite{Du:2018gyn} & \\
\hline 			   
Parameters     & $\mu=0.648$  & $\mu=0.716$   & $q_{max}=0.931$    & $\mu=1.0$     & $q_{max}=1.0$     & $q_{max}=1.0$   \\
\hline 
$a_0(980)$     & $1.0598+0.024i$     & $1.0419+0.0345i$    & $1.0029+0.0567i$  &  $\cdots$                & $\cdots$          & $\cdots$  \\
$f_0(980)$     & $0.9912+0.003i$      & $\cdots$                   & $0.9912+0.0135i$  &  $\cdots$                & $\cdots$           & $\cdots$  \\
$a_0(1710)$   & $1.7981+0.0018i$    & $1.7936+0.0094i$   & $\cdots$                  &  $1.780-0.066i$     & $1.72-0.010i$   & $1.76 \pm 0.03i$     \\
$f_0(1710)$    & $1.7676+0.0093i$   & $\cdots$                   & $\cdots$                  &  $1.726-0.014i$	     & $\cdots$           & $\cdots$  \\
\hline\hline
\end{tabular}}
\end{table}

Besides, we also calculate the ratios of branching fractions for the corresponding decay channels based on our fitting results. By integrating the $K^0_S K^0_S$ and $K^0_S \pi^+$ invariant mass distributions, we can get the ratios as below,
\begin{equation}
\begin{aligned}
\frac{\mathcal{B}(D_{s}^{+}\rightarrow S(980)\pi^{+},S(980)\rightarrow K_{S}^{0}K_{S}^{0})}{\mathcal{B}(D_{s}^{+}\rightarrow K_{s}^{0}K^{*}(892)^{+},K^{*}(892)^{+}\rightarrow K_{S}^{0}\pi^{+})} &=0.122_{-0.023}^{+0.032}, \\
\frac{\mathcal{B}(D_{s}^{+}\rightarrow S(1710)\pi^{+},S(1710)\rightarrow K_{S}^{0}K_{S}^{0})}{\mathcal{B}(D_{s}^{+}\rightarrow K_{s}^{0}K^{*}(892)^{+},K^{*}(892)^{+}\rightarrow K_{S}^{0}\pi^{+})} &=0.552_{-0.297}^{+0.460},
\label{eq28}
\end{aligned}
\end{equation}
where we make a cut at $1.5\gev$ for the lower contributions of the $S(1710)$. 
Therefore, for the $D^+_s \rightarrow S(1710) \pi^+$ decay channel, the integration limits are chosen from 1.5 to 1.77\gev, and the uncertainties are due to the limits $1.5 \pm 0.05$ and $1.77 \pm 0.05$\gev. 
Then the integration limits for the $D^+_s \rightarrow S(980) \pi^+$ processes are chosen from the corresponding threshold to 1.1\gev, and for the $D^+_s \rightarrow K^*(892)^+ K^0_S$ processes the limits are taken from $0.8\gev$ to 1.1\gev.
The uncertainties come from the changes of upper limits, which for both processes are around $1.1 \pm 0.05$\gev.
Thus, taking the branching fraction $\mathcal{B}( D^+_s \rightarrow K^{*}(892)^{+}K_{s}^{0} \rightarrow K^0_S K^0_S \pi^+)= (3.0 \pm 0.3 \pm 0.1) \times 10^{-3}$ measured by the BESIII Collaboration~\cite{BESIII:2021anf} as the input, we evaluate the branching fractions for other processes, 
\begin{equation}
\begin{aligned}
\mathcal{B}(D_{s}^{+}\rightarrow S(980)\pi^{+},S(980)\rightarrow K_{S}^{0}K_{S}^{0}) &= (0.36\pm0.04_{-0.06}^{+0.10}) \times 10^{-3},  \\
\mathcal{B}(D_{s}^{+}\rightarrow S(1710)\pi^{+},S(1710)\rightarrow K_{S}^{0}K_{S}^{0}) &= (1.66\pm0.17_{-0.89}^{+1.38}) \times 10^{-3},
\end{aligned}
\label{eq:bran}
\end{equation}
where the first uncertainties are estimated from the errors of experimental results, and the second ones are from Eq.~(\ref{eq28}). 
Compared with the experimental results, our results of the branching fraction for the decay $(D_{s}^{+}\rightarrow S(1710)\pi^{+},S(1710)\rightarrow K_{S}^{0}K_{S}^{0})$ is about $1/2$ of the measurements~\cite{BESIII:2021anf}, $(3.1 \pm 0.3 \pm 0.1) \times 10^{-3}$. Meanwhile, we also evaluate the branching fraction of the resonance $S(980)$, of which the value is $(0.36\pm0.04_{-0.06}^{+0.10}) \times 10^{-3}$.

\section{Summary}\label{Sec4}

Inspired by an enhancement near $1.7$ GeV observed in the $K^0_S K^0_S$ mass spectrum of the decay $D^+_s \rightarrow K^0_S K^0_S \pi^+$, which was reported by the BESIII Collaboration, we study this decay process exploiting the final state interaction approach based on the coupled channel interaction. 
The resonances $S(980)$ and $S(1710)$ are dynamically generated from the $S$-wave coupled channel interactions. 
Moreover, we consider the contribution of the intermediate resonance $K^*(892)^+$ in $P$ wave, which plays a key role in shaping $K^0_S \pi^+$ mass distribution.
With the contributions from both the $S$- and $P$-waves amplitudes, we make a combined fit of two invariant mass spectra of $K^0_S K^0_S$ and $K^0_S \pi^+$, where one can determine the free parameters of our formalism and get a good description of the experimental data. 
With the obtained parameters, we calculated the pole of the states $a_0(980)$, $f_0(980)$, $a_0(1710)$ and $f_0(1710)$ in the complex Riemann sheets, which are consistent with our former results. 
Note that, the poles for the $a_0(1710)$ and $f_0(1710)$ have a difference of about 30 MeV, which shows a mixing effect to the enhancement around $1.7$ GeV of the $K^0_S K^0_S$ invariant mass distribution, but with some uncertainties due to the small widths of  the poles. 
Furthermore, we calculate the branching fractions of related decay channels, where the one for the decay $D_{s}^{+}\rightarrow S(1710)\pi^{+}\rightarrow K_{S}^{0}K_{S}^{0} \pi^{+}$ is smaller than the measurement, and a prediction of the process $D_{s}^{+}\rightarrow S(980)\pi^{+} \rightarrow K_{S}^{0}K_{S}^{0} \pi^{+}$ is made, which can be measured in future experiments.

\section*{Acknowledgments}

We acknowledge Prof. Eulogio Oset for careful reading the manuscript and useful comments, and Prof. En Wang for helpful comments. 
This work is supported by the Fundamental Research Funds for the Central Universities of Central South University under Grants No. 1053320214315, 2022ZZTS0169, and the Postgraduate Scientific Research Innovation Project of Hunan Province under No. CX20220255, 
and partly by the Natural Science Foundation of Changsha under Grant No. kq2208257, the Natural Science Foundation of Hunan province under Grant No. 2023JJ30647, the Natural Science Foundation of Guangxi province under Grant No. 2023JJA110076, 
and the National Natural Science Foundation of China under Grant No. 12365019 and No. 12275364.

\end{document}